\documentclass{appolb}
\usepackage{graphicx}

\begin{document}
\title{Production of open and hidden charm in fixed-target experiments at the LHC%
\thanks{Presented by A.Szczurek at XXXI Cracow Epiphany Conference on the Recent LHC Results, Kraków, Poland, January 13-17, 2025.}%
}
\author{Antoni Szczurek
\address{Institute of Nuclear Physics, Polish Academy of Sciences, ul. Radzikowskiego 152, PL-31-342 Krak{\'o}w, Poland}
\address{College of Mathematics and Natural Sciences, University of Rzesz\'ow, PL-35-310 Rzesz\'ow, Poland}
\\[3mm]
{Anna Cisek 
\address{College of Mathematics and Natural Sciences, University of Rzesz\'ow, PL-35-310 Rzesz\'ow, Poland} 
}
\\[3mm]
{Rafa{\l} Maciu{\l}a 
\address{Institute of Nuclear
Physics, Polish Academy of Sciences, ul. Radzikowskiego 152, PL-31-342 Krak{\'o}w, Poland} 
}
}
\maketitle

\begin{abstract}
We discuss the  production of $D$ mesons and $J/\psi$ quarkonia in
proton-nucleus collisions in the fixed-target LHCb experiment. 
We consider gluon-gluon fusion within $k_t$-factorization, 
processes initiated by intrinsic charm in the nucleon and 
perturbative recombination mechanism.
All the mechanisms seem to be necessary to describe the LHCb 
experimental data. We get an upper limit for the probability
of the large-$x$ $c \bar c$ Fock component in the nucleon, which 
is slightly less than 1 \%. The recombination mechanism allows the
description of $D^0$ and $\bar D^0$ asymmetry observed
by the LHCb collaboration.\\
We also discuss the production of $J/\psi$ quarkonia, including color
singlet mechanisms. We include $g^* g^* \to J/\psi g$ and
$g^* g^* \to \chi_c(1^+,2^+)(\to J/\psi \gamma)$ within 
$k_t$-factorization approach. Different unintegrated gluon
distributions from the literature are used. A reasonable agreement 
is achieved with some distributions from the literature.
\end{abstract}
 
\section{Introduction}

There is a reach program at the LHCb to study the production of open charm
($D$ mesons) and hidden charm (charmonia). At the LHC collider mode,
the gluon-gluon fusion is the dominant process. In the collinear 
approach, rather higher-order processes have to be included.
In the $k_t$-factorization approach, the already lowest-order approach
gives a reasonable description of the data \cite{MS2022a}
There is an interesting issue of the intrinsic charm in the nucleon
which is very difficult to predict from first principles.
While the shape of $x$-distribution was predicted, e.g. by
Brodsky and collaborators \cite{BHPS} the absolute normalization
related to probability to find $c \bar c$ component must be
obtained by analysing experimental data. One possibility is
to study large energy muon neutrinos measured, e.g. by the 
Ice Cube collaboration at Antarctica.
Recently, an interesting option to measure $\tau$ neutrinos
in the LHC collider mode was proposed \cite{MS2023}.
Here, we review our recent papers on fixed-target experiments, 
which is another option to address the issue of the intrinsic charm.
The fixed-target experiments with energy $\sqrt{s} <$ 100 GeV
have observed asymmetry in the production of $D^0$ and ${\bar D}^0$.
This observation is difficult to explain in terms of 
the intrinsic charm in the nucleon, as discussed in our recent papers.

The production of quarkonia, especially $J/\psi$, is not fully 
understood so far. In general, there are colour singlet and colour
octet mechanisms. While the colour singlet mechanism in the
$k_t$-factorization approach is under better control
\cite{CS2018}, the colour octet contribution is usually fitted to 
the experimental data.
Usually, new data require a new fit of a long-distance matrix
elements.
Very recently also $J/\psi$ was measured in fixed-target $p+A$ 
collisions. We wish to discuss the situation at the lower
energies (fixed-target experiments).
We will discuss whether the approach discussed in \cite{CS2025}
can allow the description of the fixed-target LHCb data.
Our study provides a test of unintegrated gluon distributions
at larger longitudinal momentum fraction.


%


\section{Mechanisms considered}

The presented results were obtained taking into account
several mechanisms described shortly below.

\subsection{$g + g \to c + \bar c$ mechanism}

\begin{figure}
\begin{center}
 \includegraphics[width=5.5cm]{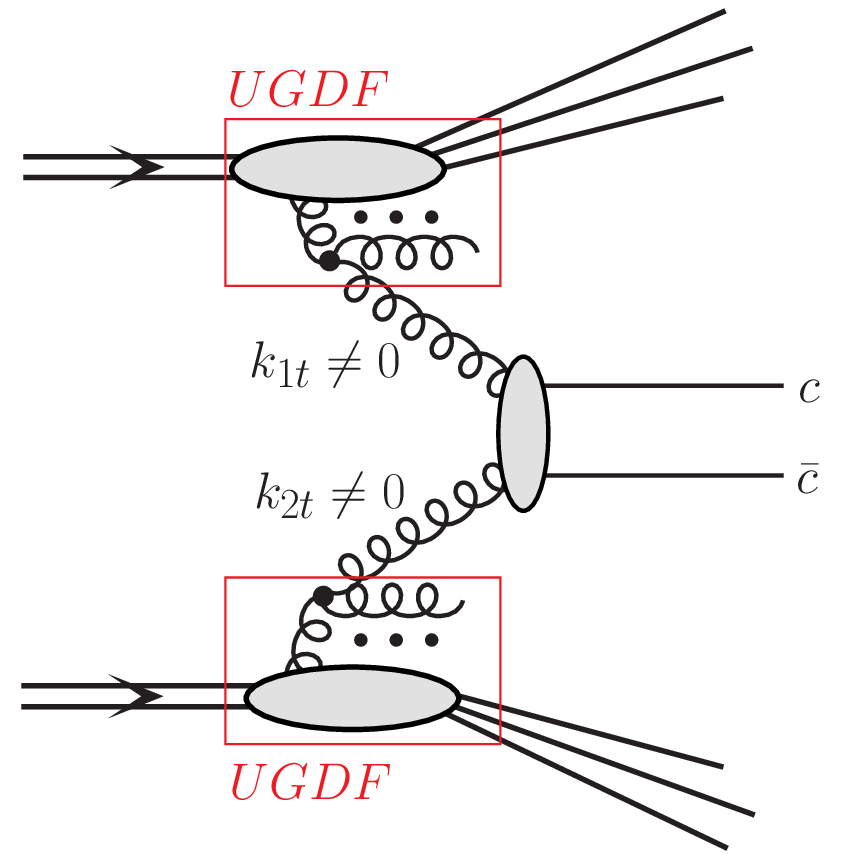}
\end{center}
\caption{Fusion of two off-shell gluons.}
\label{fig:intrinsic_charm} 
\end{figure}

At higher energies, relevant for the LHC, the dominant mechanism of 
charm/anti-charm production is gluon-gluon fusion. In the collinear approach,
one has to go to the NNLO approach to get reliable cross-sections.
The $k_t$-factorization approach is an efficient way to include 
the higher-order corrections. In this approach, the differential
cross-section can be written as:

            \begin{eqnarray}
            \frac{d \sigma}{d y_1 d y_{2} d^2 p_{1,t} d^2 p_{2,t}} =
            \int \frac{d^2 k_{1,t}}{\pi} \frac{d^2 k_{2,t}}{\pi}
            \frac{1}{16 \pi^2 (x_1 x_2 s)^2} \; 
\overline{ | {\cal M}_{g^*g^* \rightarrow Q \bar Q} |^2} 
            \;\;\;\;\;\;\;\;\;\;\;\;\;\; \nonumber \\ 
            \;\;\;\;\;\;\;\;\;\;\;\;\;\;\;\;\;\;\;\;\;\;\;\;\;\;\;\;\;\;\;\;\;\;\;\;\; \times \;\;\;
 \delta^{2} \left( \vec{k}_{1,t} + \vec{k}_{2,t} 
                 - \vec{p}_{1,t} - \vec{p}_{2,t} \right) \;
    {\cal F}_g(x_1,k_{1,t}^2,\mu) \; 
    {\cal F}_g(x_2,k_{2,t}^2,\mu) \; .   
           \end{eqnarray}

Above, 
$\overline{ | {\cal M}_{g^*g^* \rightarrow Q \bar Q} |^2}$
is the off-shell matrix element squared for $g^* g^* \to c \bar c$ 
and ${\cal F}_g(x,k_{t}^2,\mu)$ are the transverse momentum dependent,
unintegrated PDFs (uPDFs).

\subsection{Charm production driven by the 
intrinsic charm}

\begin{figure}
\begin{center}
 \includegraphics[width=5.5cm]{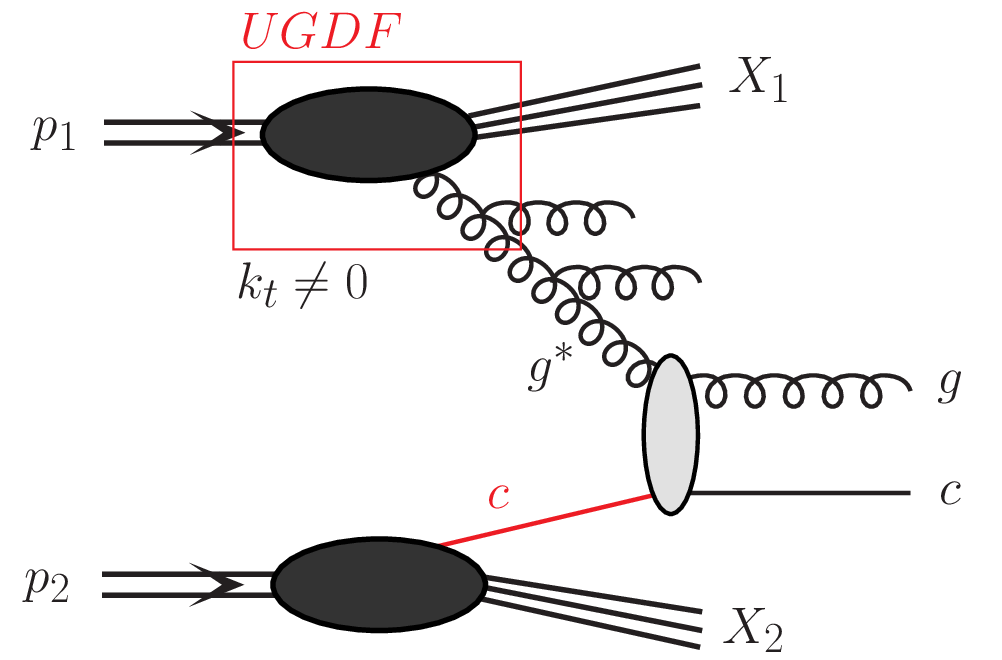}
\end{center}
\caption{A mechanism driven by the intrinsic charm.}
\label{fig:intrinsic_charm} 
\end{figure}

The differential cross section for $cg^* \to cg$ mechanism
can be written as:

\begin{eqnarray}
d \sigma_{pp \rightarrow charm}(cg^* \rightarrow c g) = \int dx_1  \int \frac{dx_2}{x_2} \int d^2k_t \, \nonumber \\ \nonumber
\times \,\, c(x_1,\mu^2) \cdot {\cal{F}}_{g} (x_2, k_t^2, \mu^2) 
\cdot d\hat{\sigma}_{cg^* \rightarrow  cg} \; .  
\end{eqnarray}
Above, $c(x_1,\mu^2)$ is the collinear charm quark PDF 
(large-$x$) while ${\cal{F}}_{g} (x_2, k_t^2, \mu^2)$
is the unintegrated gluon distribution uPDF relevant at small-$x$.

\subsection{Recombination mechanism}
 
\begin{figure}
\begin{center}
\includegraphics[width=5.5cm]{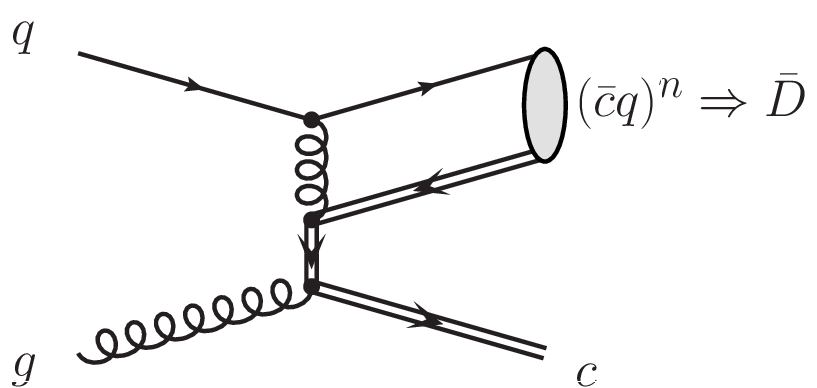} 
\end{center}
\caption{A sketch of the recombination mechanism.}
\label{fig:recombination}
\end{figure}

In our recent studies, we considered
the Braaten-Jia-Mechen (BJM) perturbative model \cite{BJM} of $D$ meson
production as illustrated in Fig.\ref{fig:recombination}.

In this approach first $q+g \rightarrow (\bar{c}q)^{n} + c$ reaction
is considered, where 
$q$ and $\bar{c}$ are in a state with definite color and angular 
momentum quantum numbers specified by $n$ which leads to
a subsequent production of a given $D$ meson.

In addition to direct $D$ meson production, we have to consider 
also fragmentation of the associated $c$-quark.

In the leading-order collinear approach the differential cross-section 
for the $qg \to \bar{D}c$ mechanism can be written as:

\begin{eqnarray}
\frac{d\sigma}{d y_1 d y_2 d^2 p_{t}} &=&
 \frac{1}{16 \pi^2 {\hat s}^2}
 [ x_1 q_1(x_1,\mu^2) \, x_2 g_2(x_2,\mu^2)
\overline{ | {\cal M}_{q g \to \bar{D} c}(s,t,u)|^2} \nonumber \\
&+& x_1 g_1(x_1,\mu^2) \, x_2 q_2(x_2,\mu^2)
\overline{ | {\cal M}_{g q \to \bar{D} c}(s,t,u)|^2} ]  \, \nonumber
\label{cross_section}
\end{eqnarray}

Above:
\begin{equation}
\overline{ | {\cal M}_{q g \to D c}(s,t,u)|^2} =
\overline{ | {\cal M}_{q g \to ({\bar c} q)^n c} |^2}
\cdot \rho.
\end{equation}

Explicit form of the matrix element squared
$\overline{ | {\cal M}_{q g \to ({\bar c} q)^n c} |^2}$ 
are available in \cite{BJM}.
$\rho$ above can be interpreted as a probability to form real meson
$D$ and can be extracted by confronting theoretical results
with experimental data.

\section{Selected results for $D$ meson production}

In \cite{MS2022a,MS2022b} and \cite{GMS2024} we presented 
many detailed results.
Here we show only some selected results from \cite{GMS2024}.

\begin{figure}
 \includegraphics[width=5.0cm]{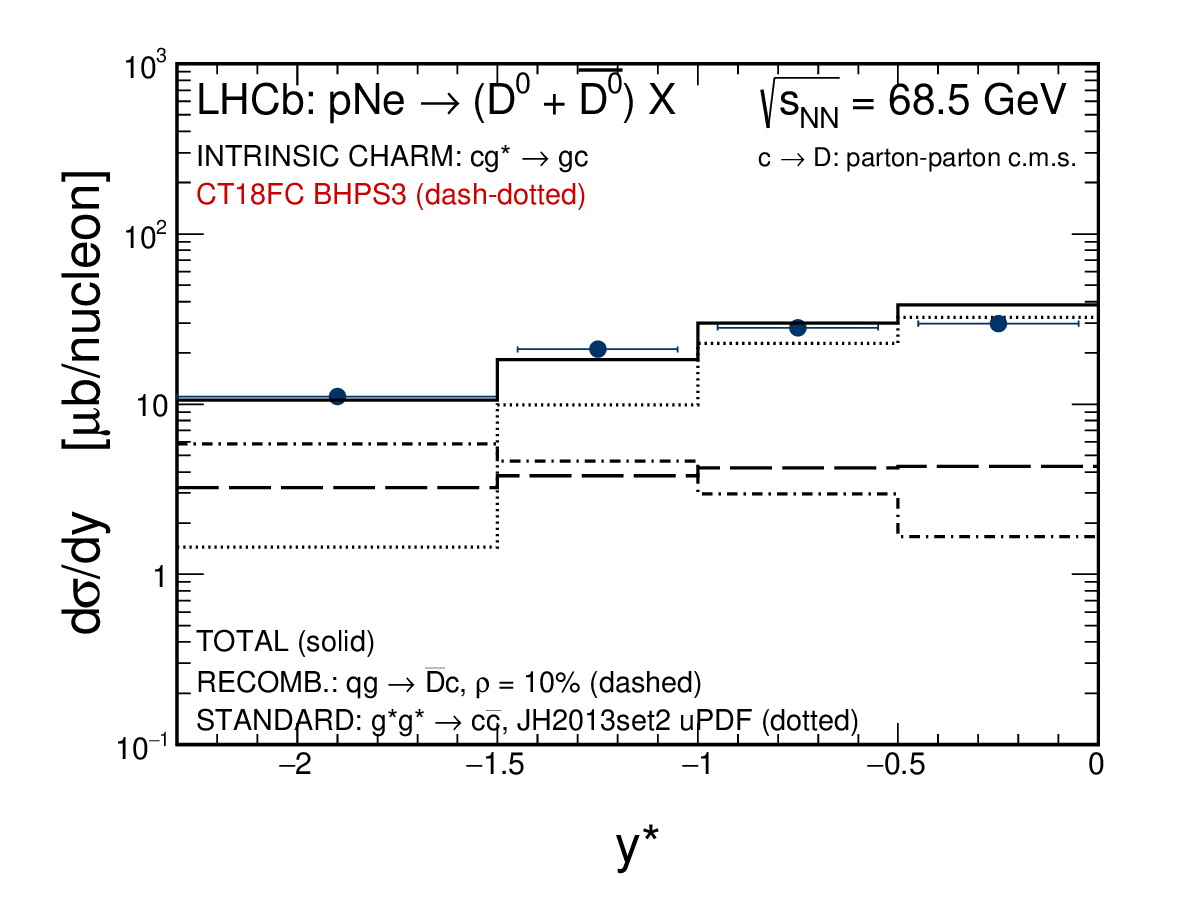} 
 \includegraphics[width=5.0cm]{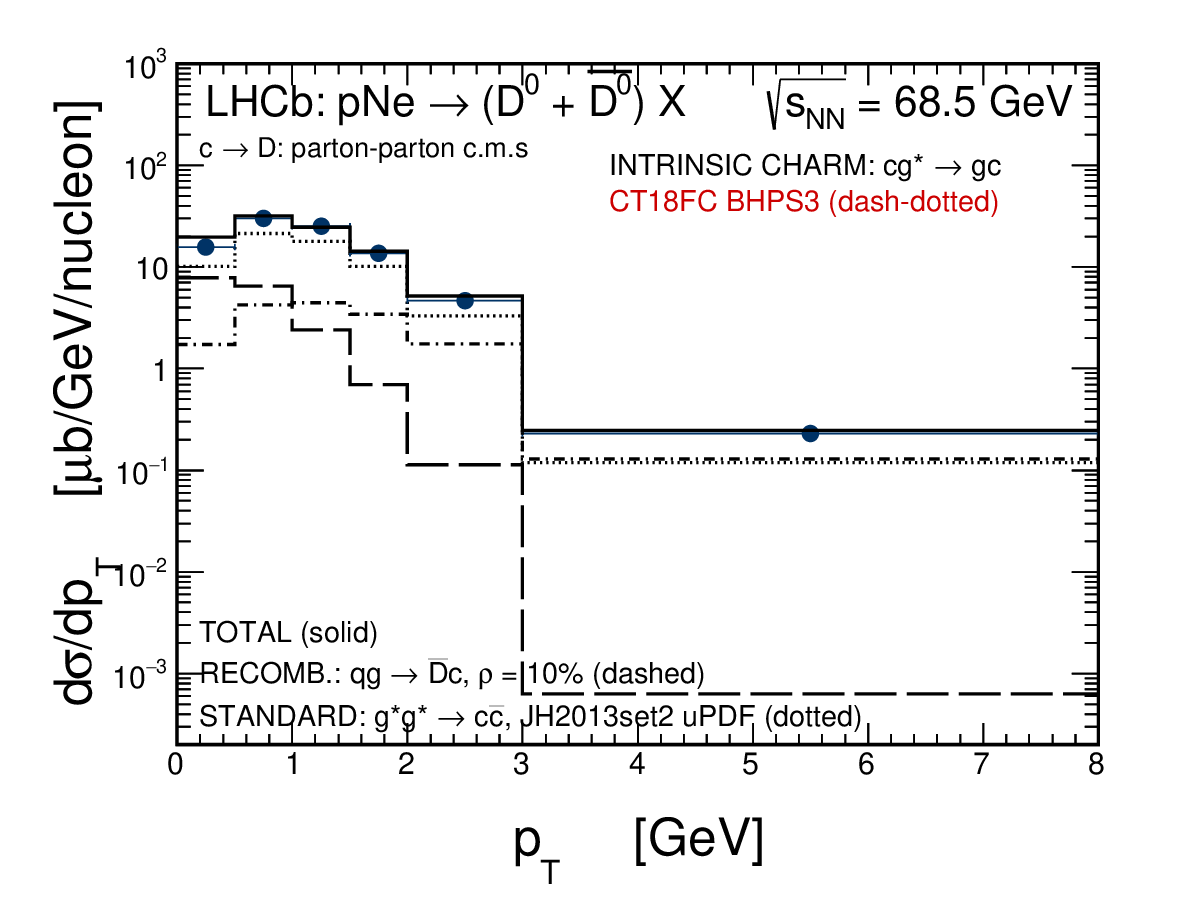} 
\caption{Rapidity and transverse momentum distributions of $D^0+{\bar D}^0$
         for BHPS model of intrinsic charm.}
\label{fig:Dmesons_BHPS}
\end{figure}

\begin{figure}
 \includegraphics[width=5.0cm]{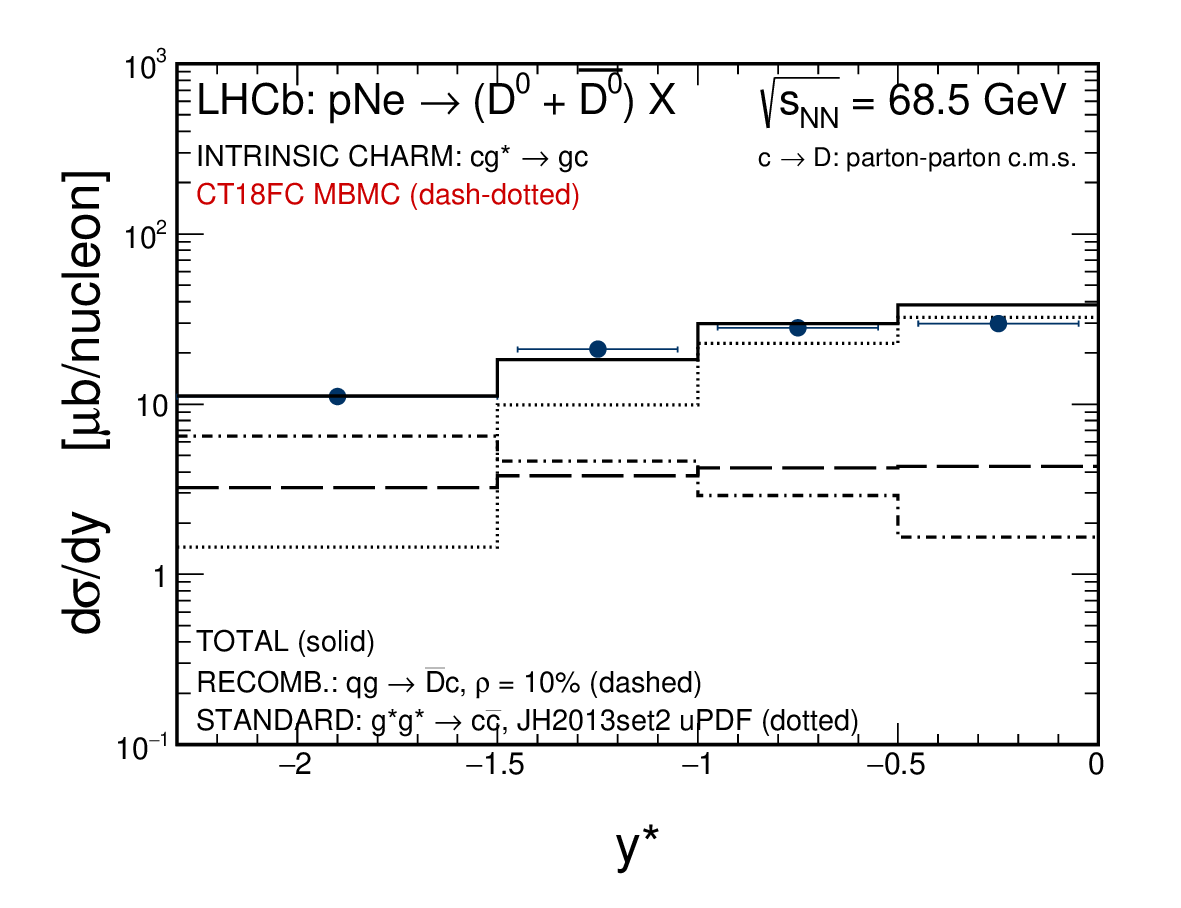}
 \includegraphics[width=5.0cm]{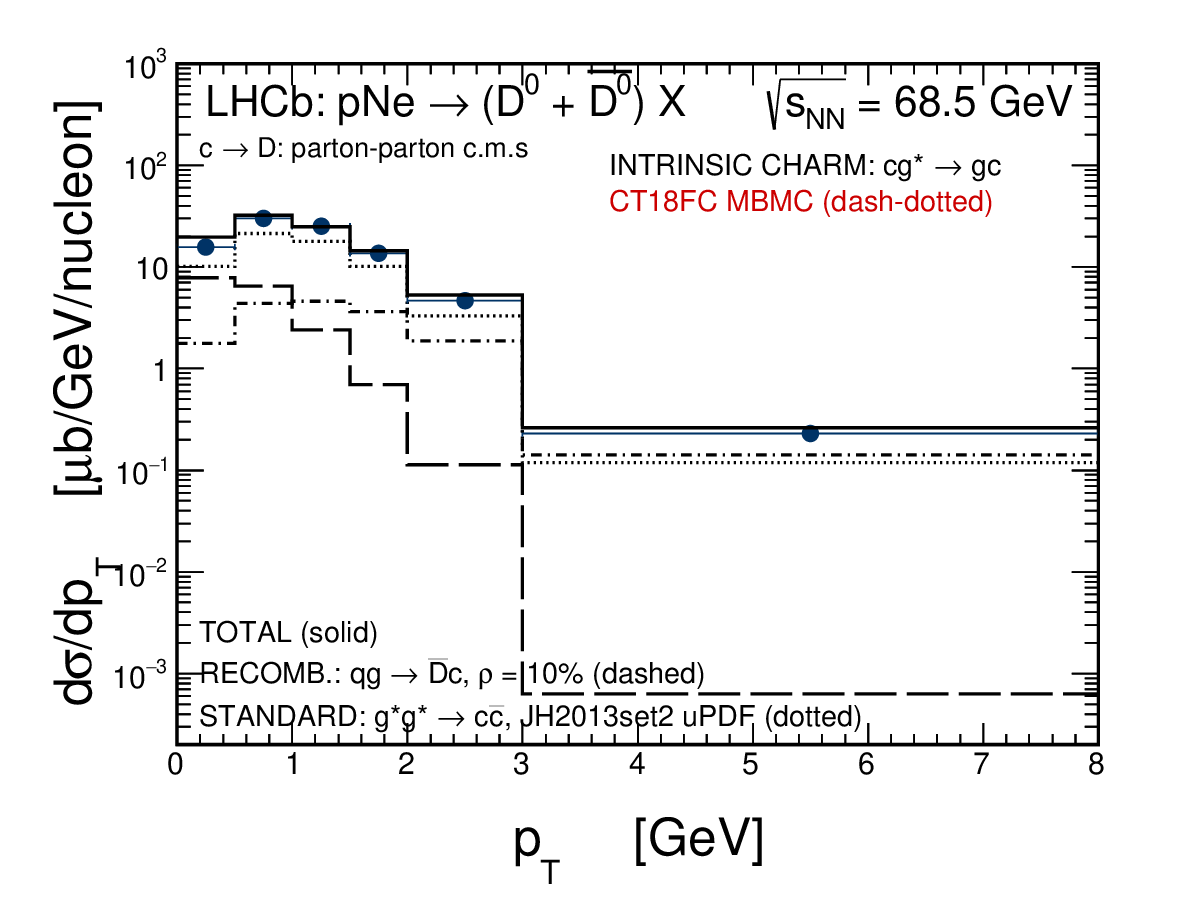} 
\caption{Rapidity and transverse momentum distribution of $D^0+{\bar D}^0$
         for the meson cloud model for the intrinsic charm.}
\label{fig:Dmesons_MBM}
\end{figure}

In Fig.\ref{fig:Dmesons_BHPS} and Fig.\ref{fig:Dmesons_MBM} we show
rapidity and transverse momentum distributions of $D^0+{\bar D}^0$
mesons for two different models of intrinsic charm: BHPS
(Brodsky-Hoyer-Peterson-Sakata) and MBM (meson-baryon model).
The results from the two models of intrinsic charm are rather similar
so it may be rather difficult to answer the question which model is
preferred by the SMOG LHCb data. The inclusion of the intrinsic charm-initiated contributions improves the description of the fixed-target
experimental data. The normalization of the intrinsic charm component
($\sim$ \%) here is adjusted to the data.

We wish to show also asymmetry defined as:
\begin{equation} 
A = \frac{\sigma_{D^0} - \sigma_{{\bar D}^0}}{\sigma_{D^0} + \sigma_{{\bar
    D}^0}} \; .
\end{equation}
We describe the experimental asymmetry with the canonical value $\rho \approx$ 0.1.

\begin{figure}
\includegraphics[width=5cm]{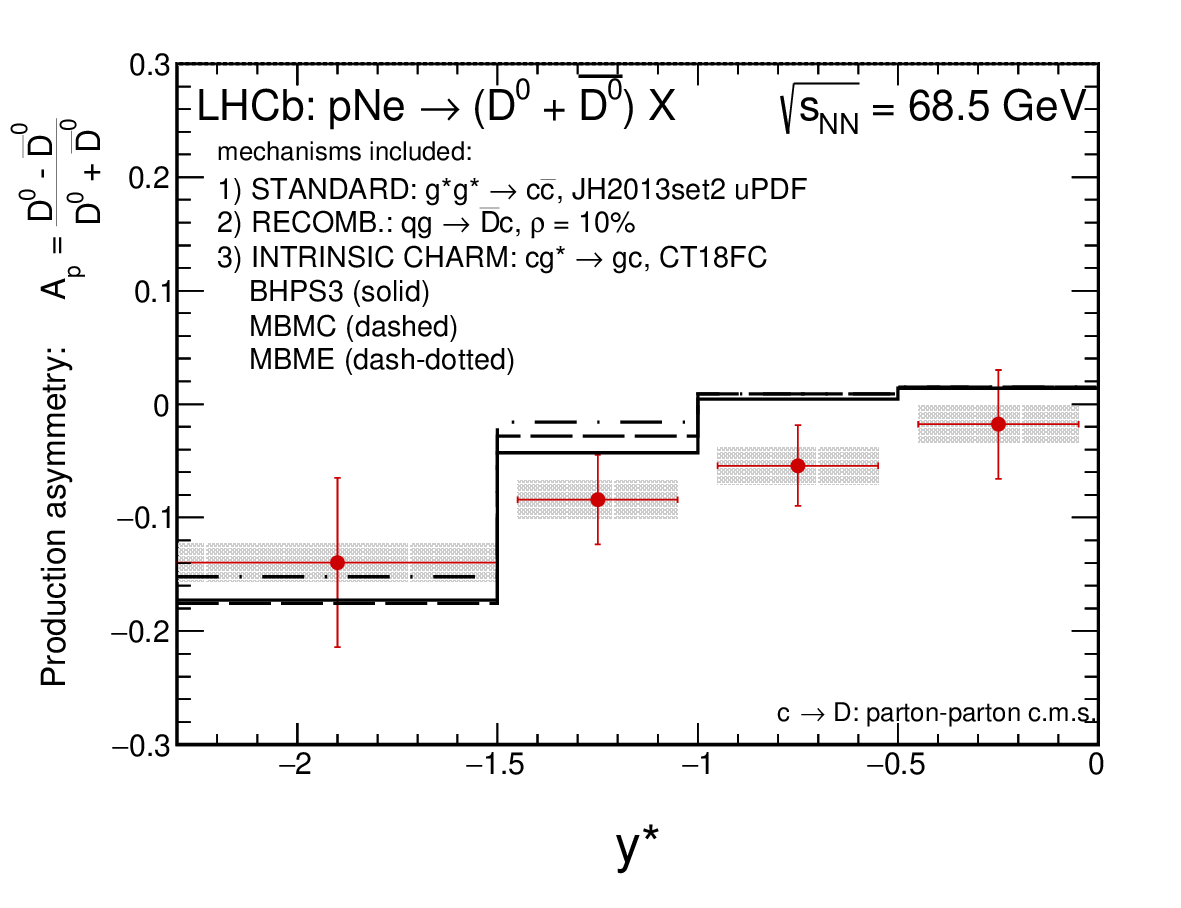}
\includegraphics[width=5cm]{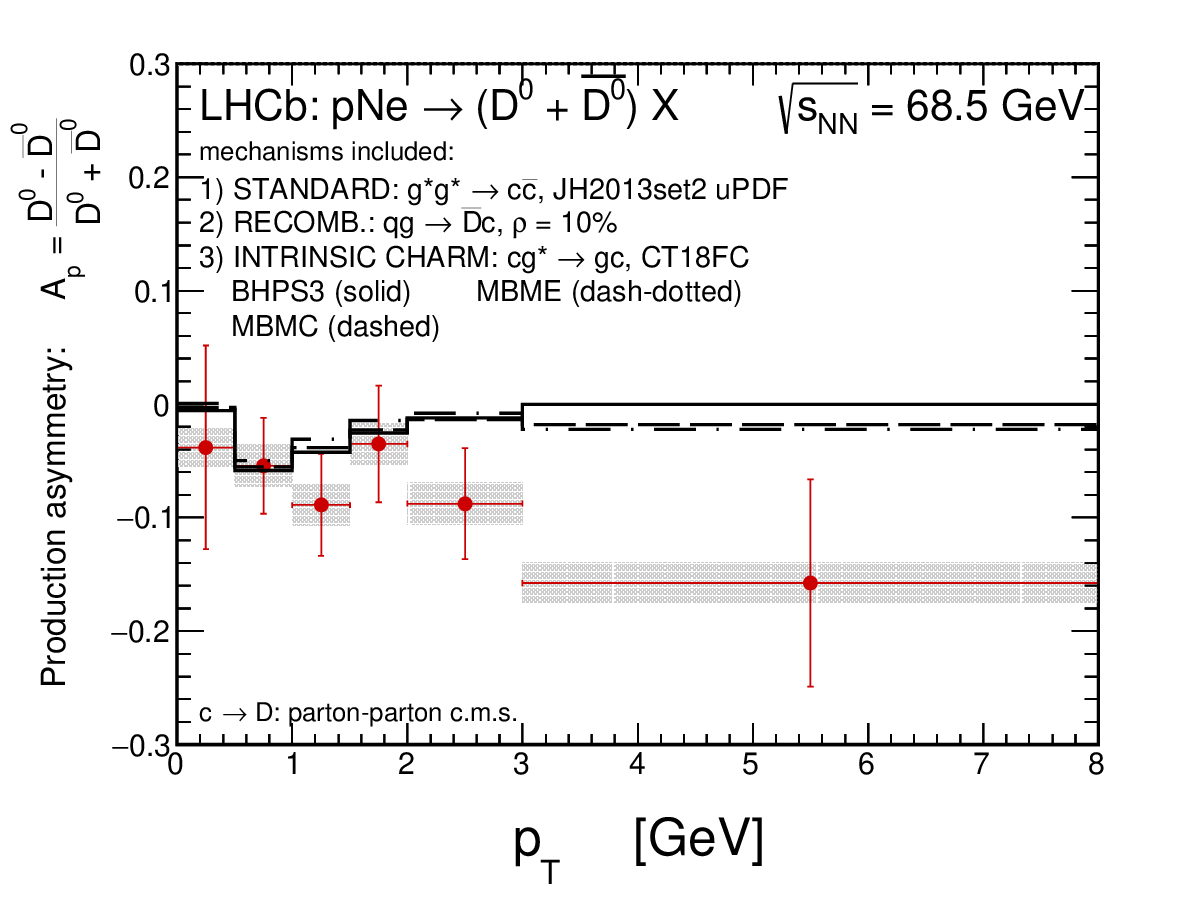}
\caption{Asymmetry as a function of $D$ meson rapidity (left) and
  transverse momentum (right).}
\label{fig:asymmetry}
\end{figure}

In summary, we can well describe the fixed-target LHCb data with
$P \approx$ 1 \% and $\rho \approx$ 0.1.
The value of the probability found here is almost the same
as found in other processes (see e.g.\cite{GMS2022}).

\section{$p p \to J/\psi$}

The production of $J/\psi$ in proton-proton collisions is known as 
a rather challenging task.
Different authors use different long-distance 
matrix elements to get a satisfactory description of the data.
Here we shall concentrate on prompt $J/\psi$ production,
 i.e. the decays of $B, \bar B$ will be neglected.
The direct production is not sufficient and one has to include also
decays of other quarkonia which give a sizable contribution.

Some time ago we showed that the $k_t$-factorization approach
with non-relativistic approximation and unintegrated gluon distributions
provides quite a good description of the world data \cite{CS2018}.
This is because the $k_t$-factorization approach 
includes effectively higher-order corrections.
In the present studies, we wish to test how good is such an approach
at lower energies.
A few years ago the LHCb collaboration presented the first result 
for fixed-target experiments using the so-called 
SMOG device.
So far the unintegrated UGDFs have been tested in different processes
rather at energies in which one is sensitive to the region
of not too high longitudinal momentum fraction carried by gluon
in the proton (nucleon).
The region of UGDFs at larger values of $x$ was not well 
tested so far.
Therefore the relatively new fixed-target data give a chance for such
tests.

We calculate the dominant color-singlet $g g \to J/\psi g$
contribution taking into account the transverse momenta of initial gluons.

\begin{figure}
\begin{center}
\includegraphics[width=5cm]{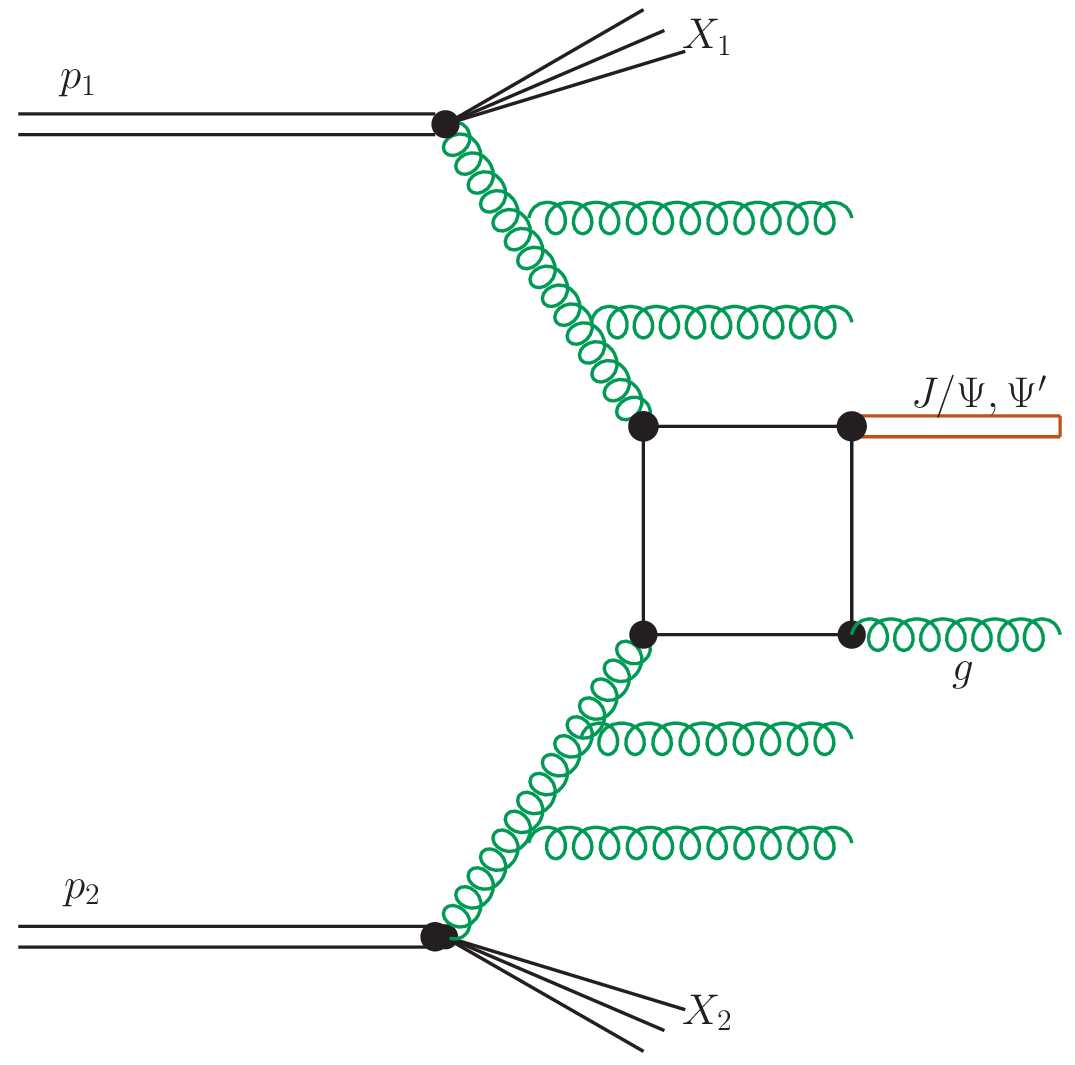}
\end{center}
\caption{The diagram for direct $J/\psi$ ($\psi'$) meson production
in the $k_t$-factorization approach.}
\label{fig:gg_Jpsig}
\end{figure}

In the $k_t$-factorization approach the differential cross-section can 
be written as:
\begin{eqnarray}
&&\frac{d \sigma(p p \to J/\psi g X)}{d y_{J/\psi} d y_g d^2 p_{J/\psi,t} d^2 p_{g,t}}
 = 
\frac{1}{16 \pi^2 {\hat s}^2} \int \frac{d^2 q_{1t}}{\pi} 
\frac{d^2 q_{2t}}{\pi} 
\overline{|{\cal M}_{g^{*} g^{*} \rightarrow J/\psi g}^{off-shell}|^2} 
\nonumber \\
&& \times \;\; 
\delta^2 \left( \vec{q}_{1t} + \vec{q}_{2t} - \vec{p}_{H,t} -
  \vec{p}_{g,t} \right)
{\cal F}_g(x_1,q_{1t}^2,\mu^2) {\cal F}_g(x_2,q_{2t}^2,\mu^2)  ;
\label{kt_fact_gg_jpsig}
\end{eqnarray}
where ${\cal F}_g$ are the unintegrated gluon distributions functions.

The corresponding matrix element squared for the 
$g g \to J/\psi g$ is
\begin{equation}
|{\cal M}_{gg \to J/\psi g}|^2 \propto \alpha_s^3 |R(0)|^2 \; .
\label{matrix_element} 
\end{equation}
%

\begin{figure}
\begin{center}
\includegraphics[width=5cm]{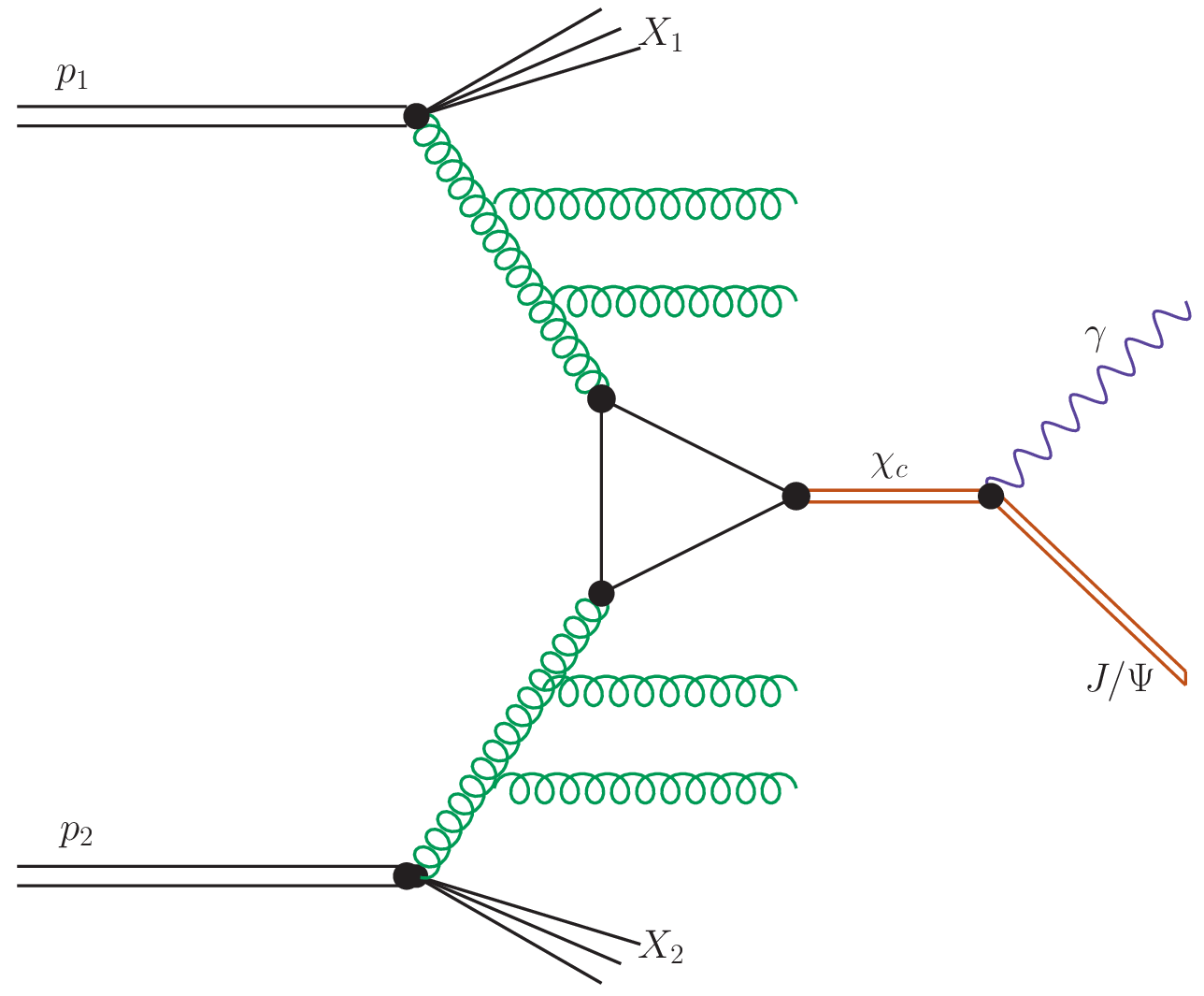}
\end{center}
\caption{The diagram for direct $J/\psi$ ($\psi'$) meson production
in the $k_t$-factorization approach.}
\label{fig:gg_Jpsig}
\end{figure}

In the $k_t$-factorization approach the leading-order cross-section 
for the $\chi_c$ meson production can be written somewhat formally as:
\begin{eqnarray}
\sigma_{pp \to \chi_c} &=& \int \frac{dx_1}{x_1} \frac{dx_2}{x_2}
\frac{d^2 q_{1t}}{\pi} \frac{d^2 q_{2t}}{\pi} 
\delta \left( (q_1 + q_2)^2 - M_{\chi_c}^2 \right) 
\sigma_{gg \to H}(x_1,x_2,q_{1},q_{2}) \nonumber \\
&&\times \; 
{\cal F}_g(x_1,q_{1t}^2,\mu_F^2) {\cal F}_g(x_2,q_{2t}^2,\mu_F^2)
\; ,
\label{chic_kt_factorization}
\end{eqnarray}
where ${\cal F}_g$ are the unintegrated (or transverse-momentum-dependent) 
gluon distributions and $\sigma_{g g \to \chi_c}$ is 
$g g \to \chi_c$ (off-shell) cross section.

In our recent analysis \cite{CS2025} we made calculations with 
the following unintegrated gluon distribution functions used 
previously in the literature:

\begin{itemize} 
\item (a) Kimber-Martin-Ryskin (KMR),
\item (b) Jung-Hautmann (JH2013), 
\item (c) Gaussian with $\sigma = 0.5$ GeV and CTEQ-Tea Parton 
          Distribution Functions,
\item (d) Kharzeev-Levin (KL), 
\item (e) Kutak-Stasto (KS), 
\item (f) Moriggi-Peccini-Machado (MPM).
\end{itemize}

\begin{figure}[h]
\centering
\includegraphics[width=4.0cm]{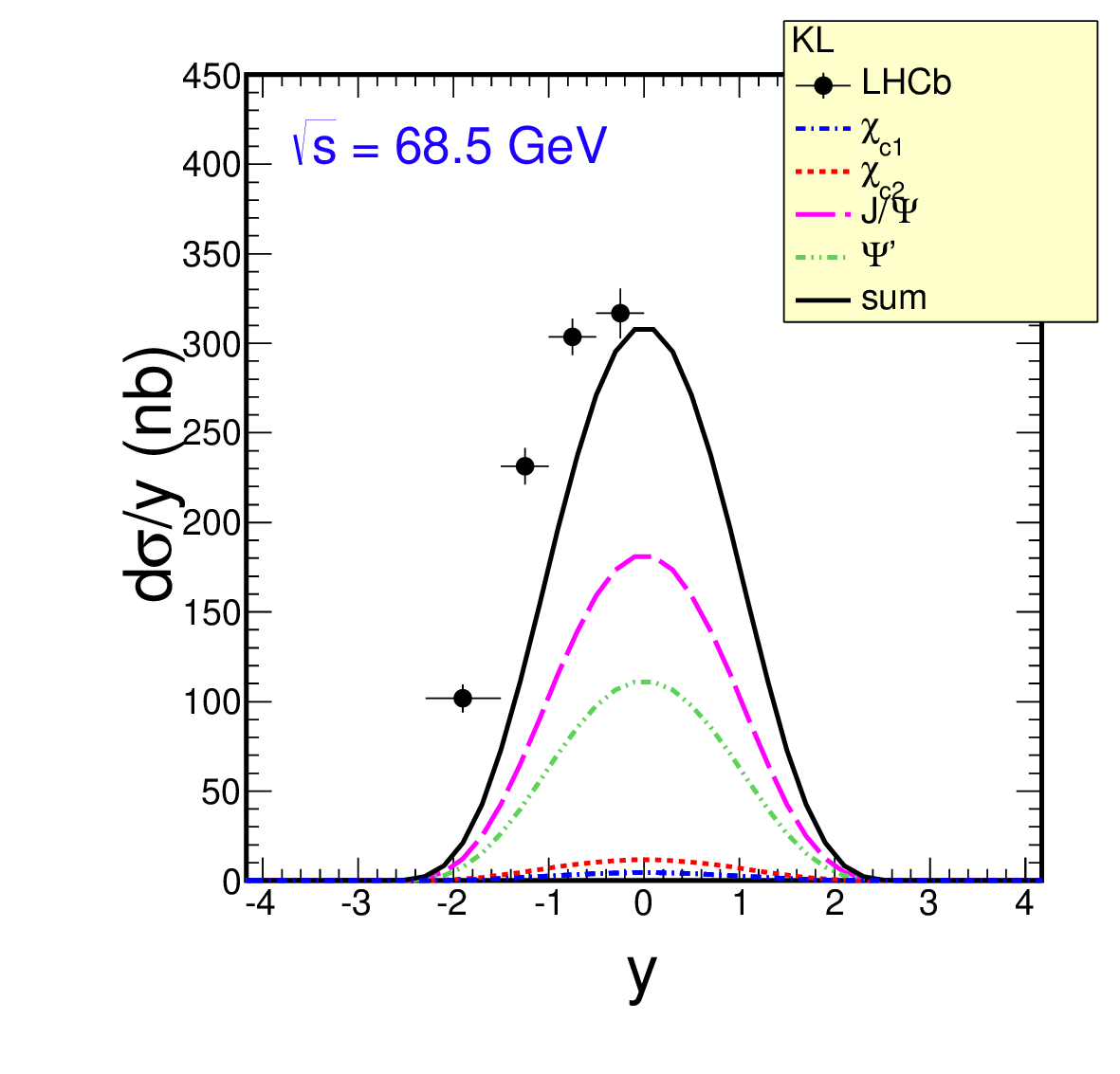}
\includegraphics[width=4.0cm]{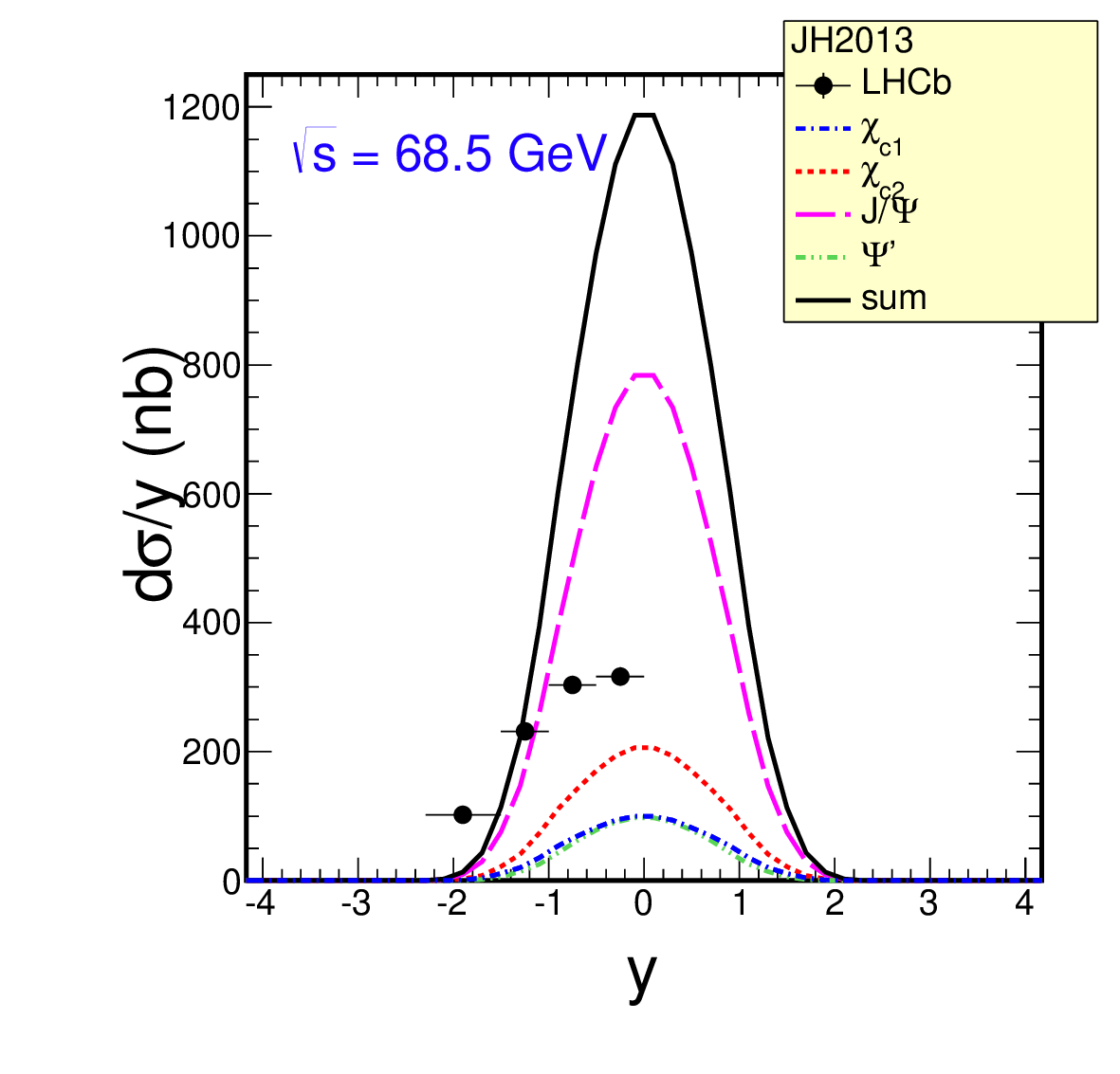}\\
\includegraphics[width=4.0cm]{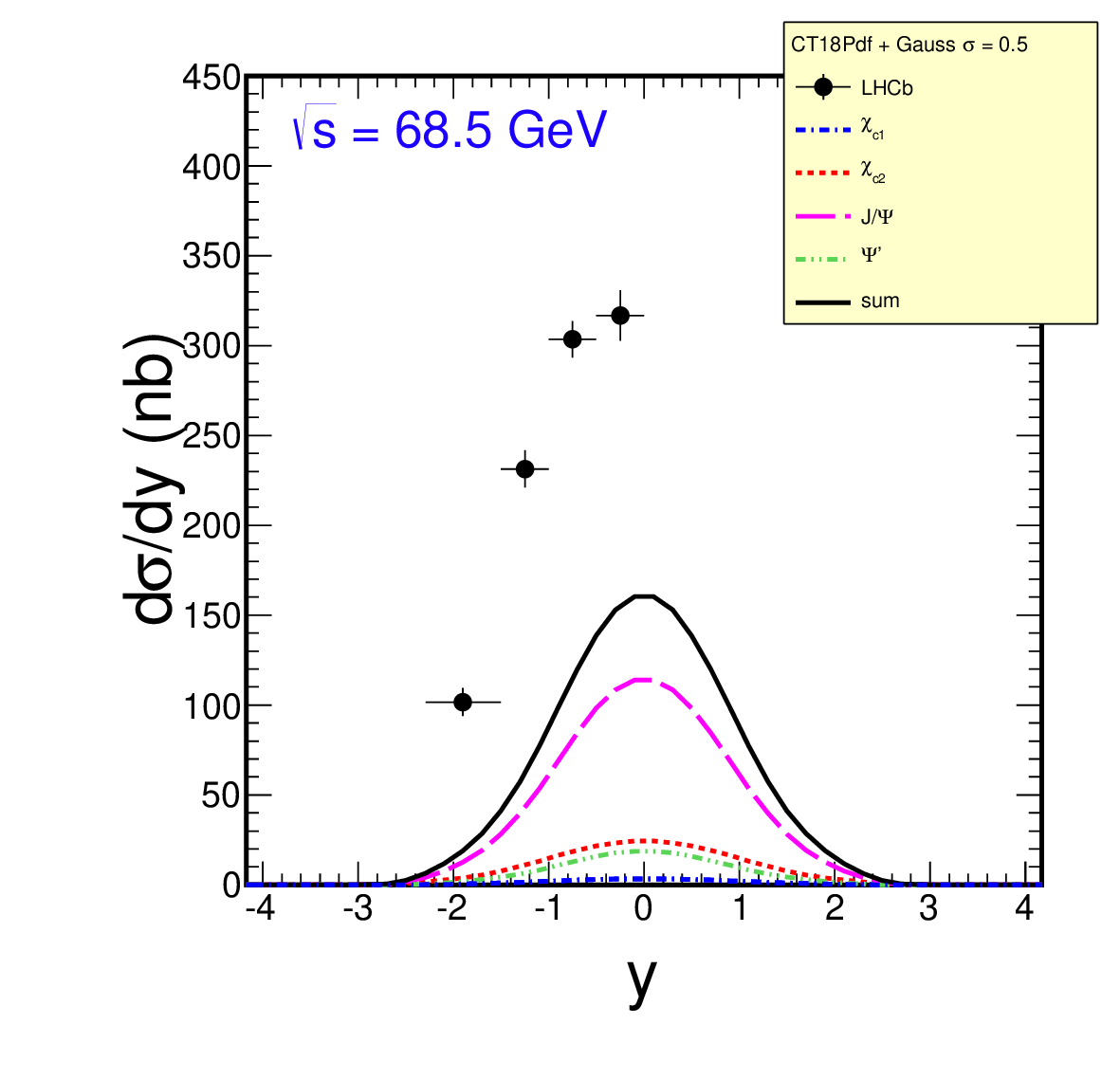}
\includegraphics[width=4.0cm]{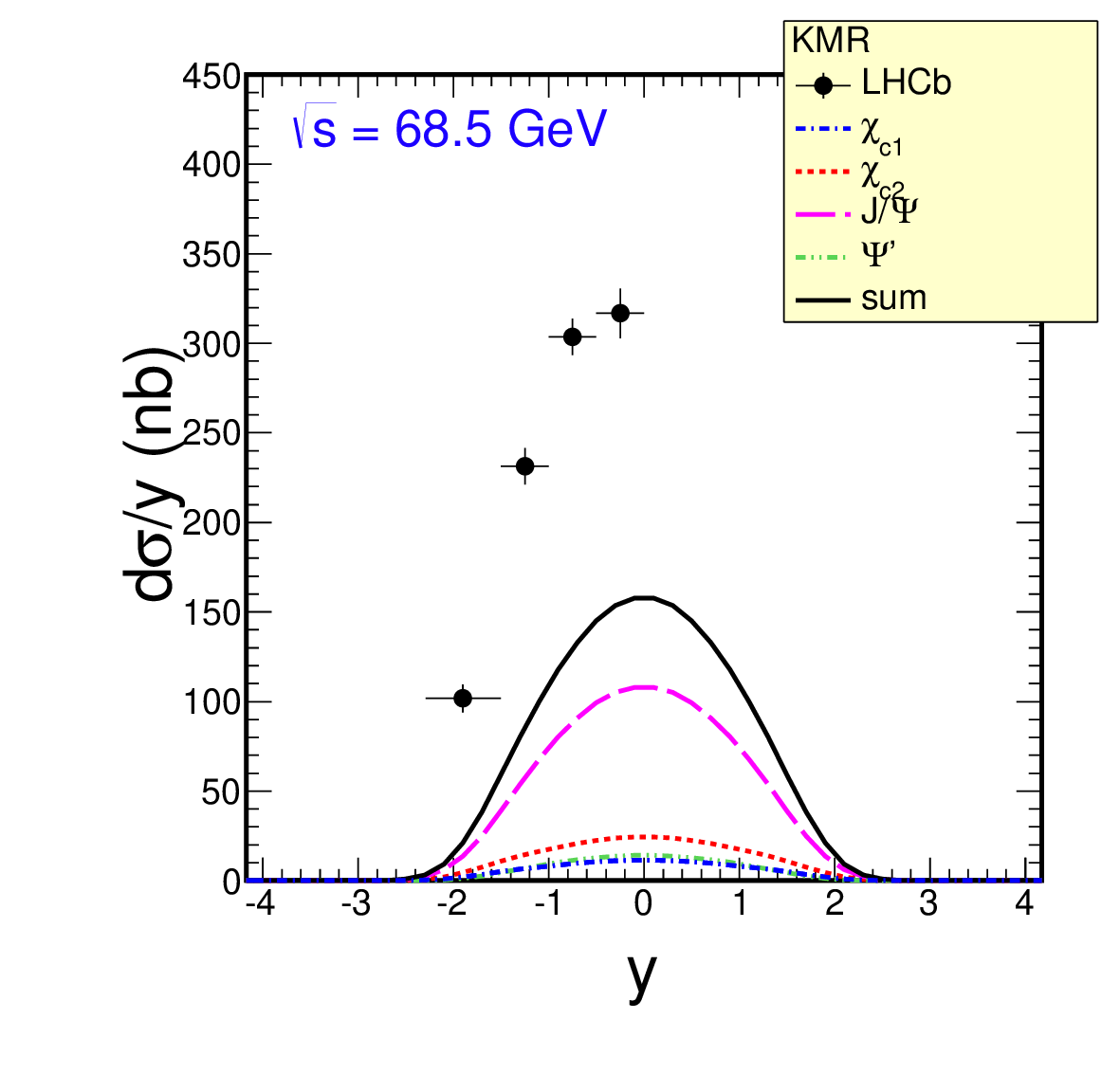}
\caption{Rapidity distribution of $J/\psi$ mesons for all considered mechanisms
for different unintegrated distribution functions.}
\label{fig_rapidity}
\end{figure}

\begin{figure}[h]
\centering
\includegraphics[width=4.0cm]{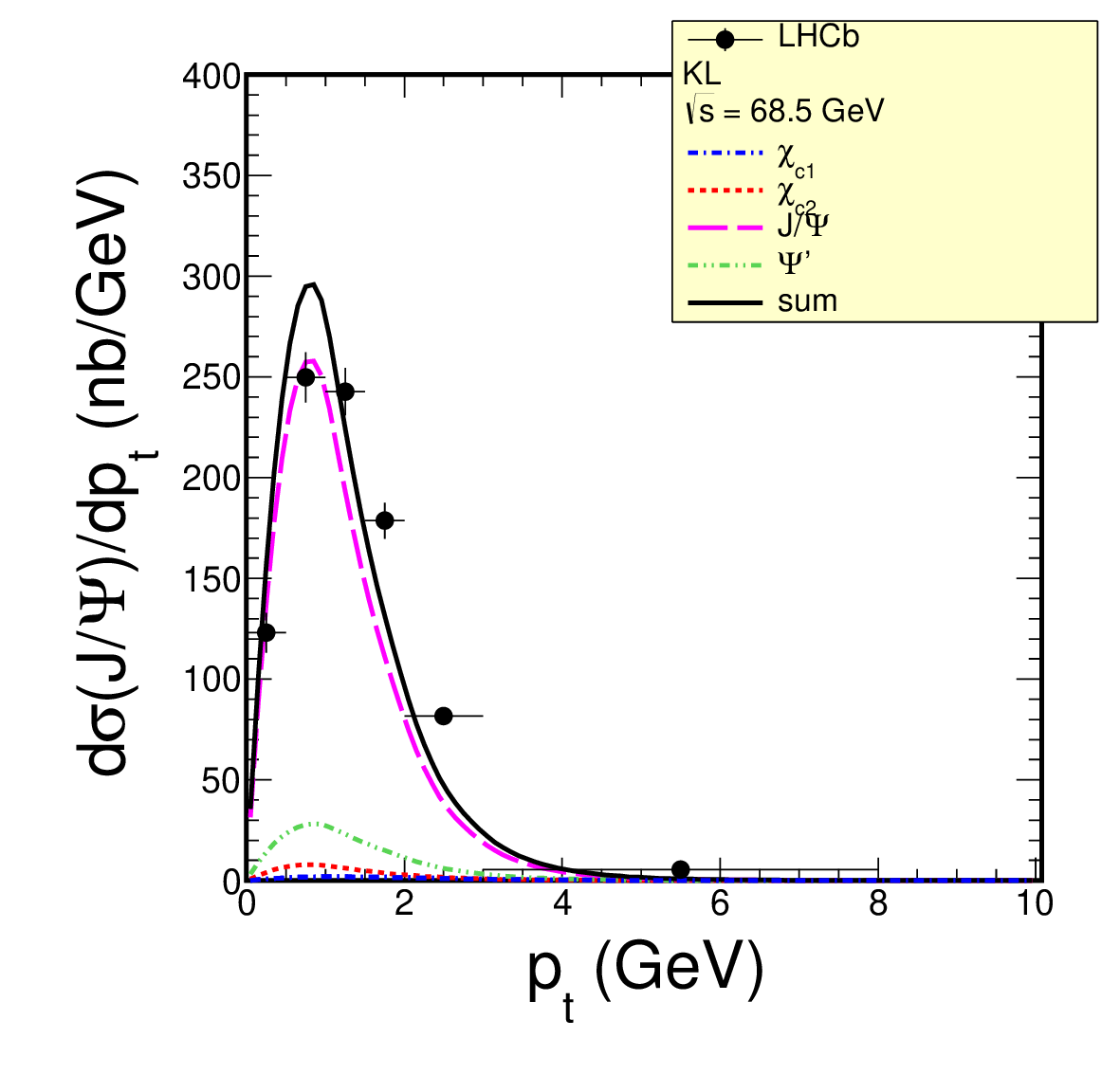}
\includegraphics[width=4.0cm]{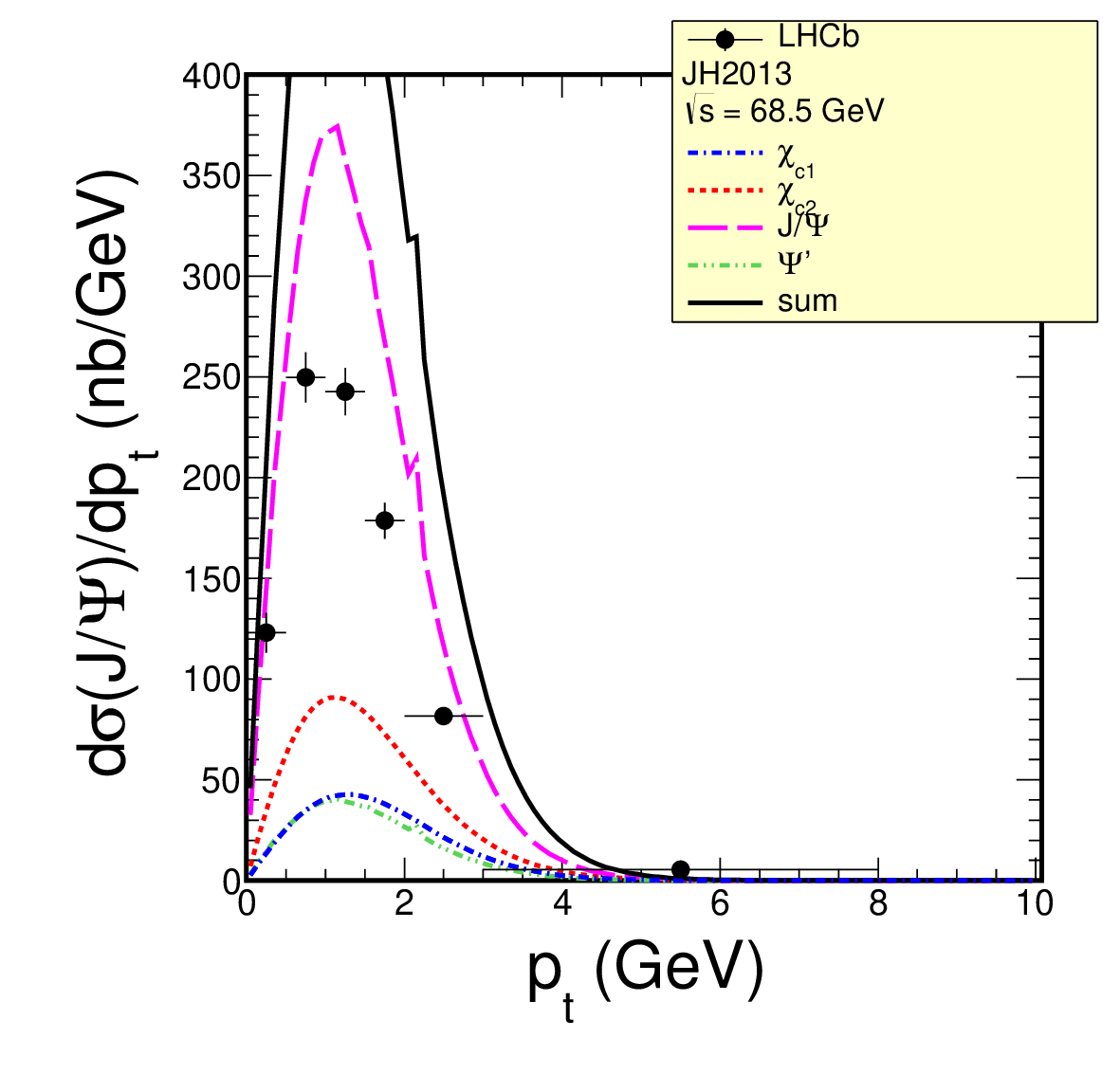}\\
\includegraphics[width=4.0cm]{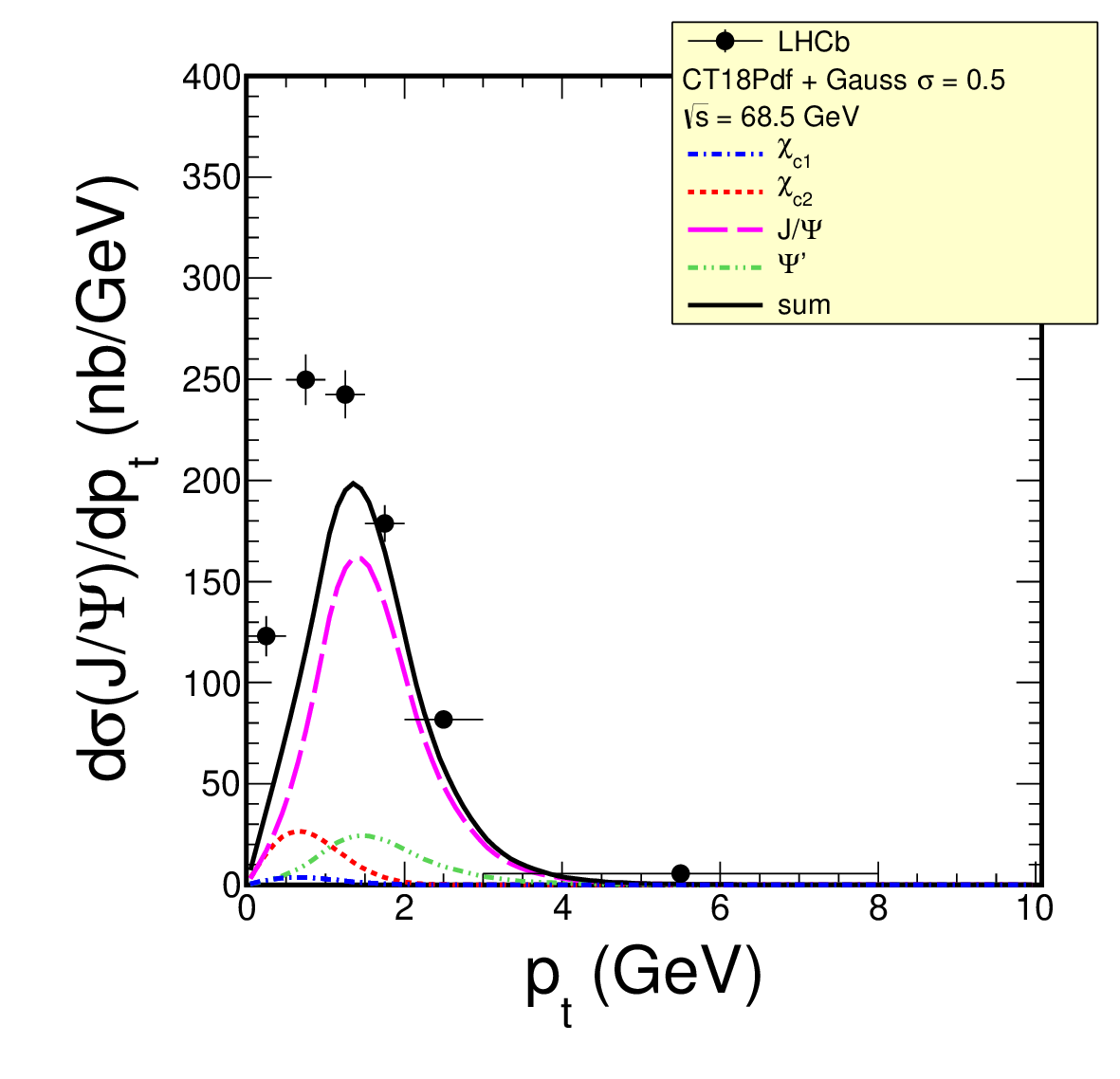}
\includegraphics[width=4.0cm]{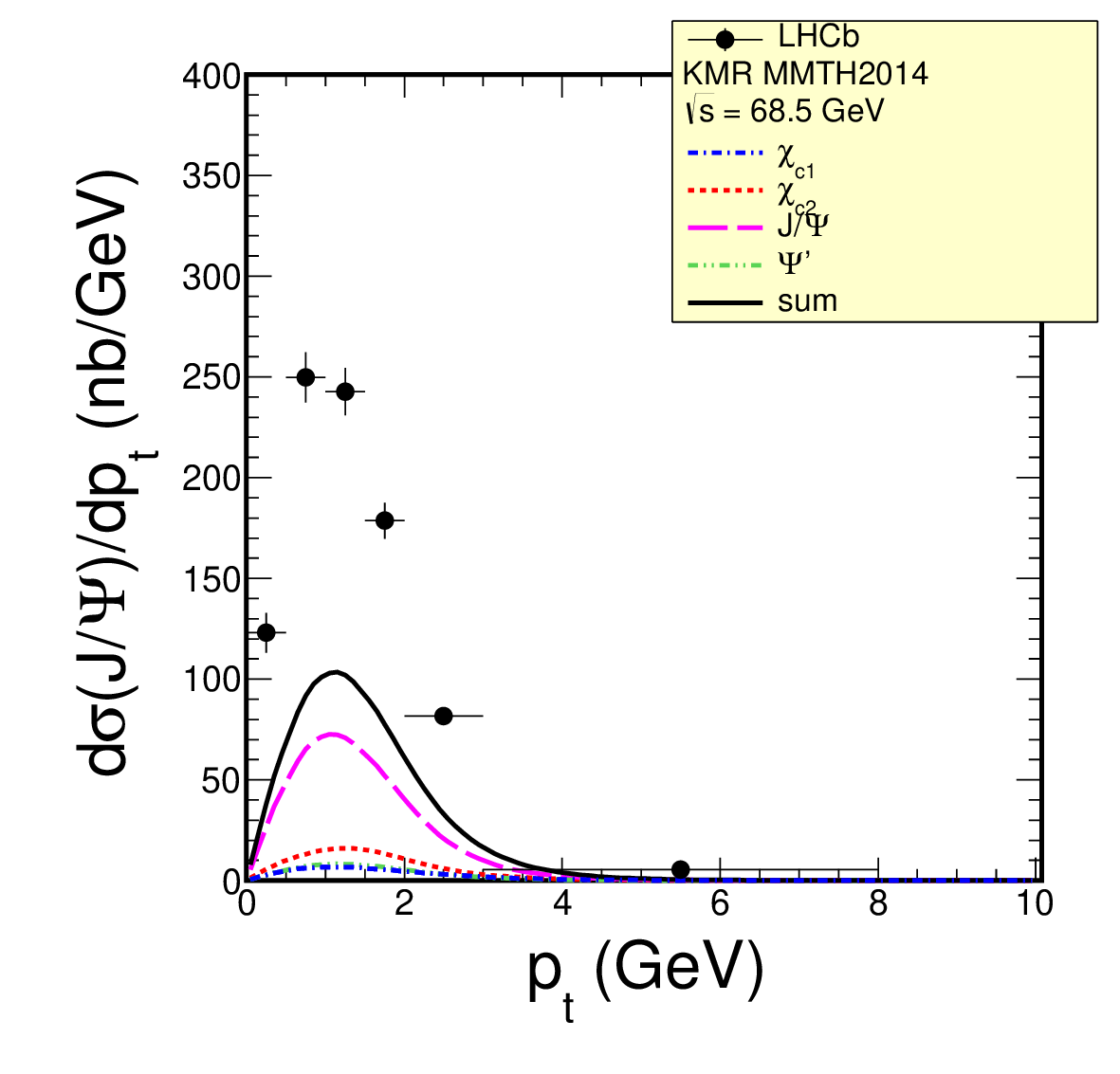}
\caption{Transverse momentum distribution of $J/\psi$ mesons for all considered mechanisms
for different unintegrated distribution functions.}
\label{fig_pt}
\end{figure}

For some UGDFs the cross-section is too small, for others it is much 
too large, but for some of them the results are close to the 
experimental data.

\section{Conclusions}

We have shown that the intrinsic charm and the recombination 
mechanisms are extremely important for forward charm production 
at energies much lower than the nominal LHC energies.
A scenario proposed with the intrinsic charm contribution is needed 
to describe the data points in the backward direction and at a larger 
$p_{T}$'s.
We have found an upper limit for the intrinsic charm probability 
$P_{IC}$ ($\approx 0.5\%$) with the CT18FC.
The recombination probability from $D/\overline{D}$-production 
asymmetry ($\rho \approx$ 10\% has been found, consistent with earlier
findings in the literature.
The $D/\bar{D}$ production asymmetry in the backward region 
and at small transverse momenta has been explained by 
the recombination mechanism.
The asymmetry at larger transverse momenta can be described neither by the recombination mechanism nor by the asymmetric intrinsic 
charm.

We have analyzed also prompt production of $J/\psi$ quarkonia 
for energies corresponding to fixed-target LHCb p + A data 
at $\sqrt{s} =$ 68.5 GeV. 
In this exploratory study, we have performed
calculations within $k_t$-factorization approach as we did 
previously in high-energy collisions.
At high energies, one is sensitive to the region of very small
$x$ of the order of 10$^{-4}$ while here one is sensitive to much 
larger values of $x_1$ or $x_2$, typically larger than 10$^{-2}$.
We have calculated distributions in rapidity and 
transverse momentum.
The obtained results have been compared to the LHCb data.
There is relatively large spread of results
for this intermediate-$x$ region. The KMR and 
JH2013 UGDFs give reasonable description of the data. Quite good result 
has been obtained with the KL UPDF used previously to light charged 
particle production.


\end{document}